\documentclass[runningheads,a4paper]{llncs}
\usepackage{amssymb}
\usepackage{graphicx}
\graphicspath{ {figures/} }
\usepackage{caption}
\usepackage[caption=false]{subfig}

\usepackage{amsmath}

\usepackage{times}
\usepackage{epsfig}
\usepackage{amssymb}
\usepackage[linesnumbered,ruled,vlined]{algorithm2e}
\usepackage{xfrac}

\usepackage{url}
\urldef{\mailsa}\path|{amitboy,bron}@cs.technion.ac.il| 
\urldef{\mailsb}\path|michael.bronstein@usi.ch|
\newcommand{\keywords}[1]{\par\addvspace\baselineskip
\noindent\keywordname\enspace\ignorespaces#1}

 {\everymath{\displaystyle\everymath{}}\array}%
 {\endarray}
\everymath{\displaystyle\everymath{}}

\newcommand{\bm}[1]{\boldsymbol{#1}}

\newcommand{\bb}[1]{\bm{\mathrm{#1}}}

\def\Tr{\mathrm{T}}

\def\trace{\mathrm{tr}}

\begin{document}

\mainmatter  

\title{Subspace Least Squares Multidimensional Scaling}

\titlerunning{Subspace LS-MDS}

%
%
\author{Amit Boyarski\inst{1}%
\and Alex M. Bronstein\inst{1,2,3}\and Michael M. Bronstein\inst{2,3,4}}
\authorrunning{Subspace LS-MDS}

\institute{Technion, Israel\and Tel-Aviv University, Israel \and Intel Perceptual Computing, Israel \and University of Lugano, Switzerland\\
\mailsa\\
\mailsb\\
}
%
%
\newcommand{\samelineand}{\qquad}

\toctitle{Subspace LS-MDS}
\tocauthor{Subspace LS-MDS}
\maketitle

\begin{abstract}
  Multidimensional Scaling (MDS) is one of the most popular methods for dimensionality reduction and visualization of high dimensional data. Apart from these tasks, it also found applications in the field of geometry processing for the analysis and reconstruction of non-rigid shapes. In this regard, MDS can be thought of as a \textit{shape from metric} algorithm, consisting of finding a configuration
of points in the Euclidean space that realize, as isometrically as
possible, some given distance structure. In the present work we cast the least squares variant of MDS (LS-MDS) in the spectral domain. This uncovers a multiresolution property of distance scaling which speeds up the optimization by a significant amount, while producing comparable, and sometimes even better, embeddings.
\keywords{multidimensional scaling, SMACOF, majorization, multi-resolution, spectral regularization, canonical forms}
\end{abstract}

\section{Introduction}
Multidimensional Scaling (MDS) is one of the most popular techniques for dimensionality reduction, whose purpose is to represent dissimilarities between objects as distances between points in a low dimensional space. Since its invention in the 50's by Torgerson \cite{torgerson1952multidimensional} within the field of psychometrics, it has been used in diverse fields including sociology, marketing, and colorimetry, to name a few \cite{borg2005modern,lawrence2011user}. More recently, MDS has found applications in the field of geometry processing, in such tasks as analysis and synthesis of non-rigid shapes \cite{bronstein2008numerical,boscaini2014shape}. In this regard, MDS can be thought of as a \textit{shape from metric} algorithm, consisting of finding a configuration of points in the Euclidean space that realize, as isometrically as possible, some given distance structure. 
Least squares Multidimensional Scaling (LS-MDS) is a particular MDS model which seeks a low dimensional Euclidean embedding by minimizing a geometrically meaningful criterion, namely the \textit{Kruskal stress} 
\begin{equation}\label{eq:kruskalstress}
\sigma(\bb{X}) = \sum_{i<j}w_{ij}\left(\|\bb{x}_i-\bb{x}_j\|-d_{ij}\right)^2.
\end{equation}
Here $d_{ij}$ is a (symmetric) measure of dissimilarity between two sample points $i$ and $j$; $w_{ij}>0$ is a (symmetric) weight assigned to the pairwise term between those samples, and $\bb{X}\in\mathcal{\mathbb{R}}^{N\times m}$ are the coordinates of the low-dimensional Euclidean embedding. Since (\ref{eq:kruskalstress}) is a nonlinear non-convex function, a variety of non-linear optimization techniques have been employed for its minimization, often resulting in local minima whose quality crucially depends on the initialization. Of these techniques, one of the most popular approaches is the SMACOF algorithm discovered by Jan de-Leeuw \cite{de1977applications}, which stands for Scaling by Majorizing a Complicated Function. SMACOF is essentially a gradient based method, which suffers from the typical slow convergence associated with first order optimization methods. To supplement on this, a single iteration of SMACOF requires the computation of the Euclidean pairwise distances between all points participating in the optimization at their current configuration, a time consuming task on its own, which limits its application to small size data. 

A related family of methods with the same goal in mind, confusingly called classical (multidimensional) scaling, aims at finding a low-dimensional Euclidean embedding $\bb{X}\in\mathbb{R}^{N\times m}$ by minimizing the following algebraic criterion called \textit{strain}
\begin{equation}\label{eq:strain}
s(\bb{X}) = \|\bb{X}\bb{X}^\Tr+\frac{1}{2}\bb{H}^\Tr\bb{D}\bb{H}\|_\mathrm{F}^2 
\end{equation}
where $\bb{H} = \bb{I}-\frac{1}{n}\bb{1}\bb{1}^\Tr$ is a \textit{centering matrix} transforming a Euclidean squared-distance matrix $\bb{D}$ into a Gram matrix of inner products by fixing the origin, and $\bb{1}$ denotes the vector of ones. If $\bb{D}$ is a Euclidean squared-distance matrix between points in $\mathbb{R}^n$, $-\frac{1}{2}\bb{H}^\Tr\bb{D}\bb{H}$ is a Gram matrix of rank (at most) $n-1$, and can be faithfully approximated by a Gram matrix of lower rank. However, if $\bb{D}$ is not a Euclidean squared-distance matrix, one can only hope that it is indeed well approximated by a low-rank Gram matrix after applying the double centering transformation.

Though (\ref{eq:strain}) is not convex, it can be minimized globally by eigendecomposition techniques. On top of being expensive to apply, these techniques require the availability of the complete matrix $\bb{D}$, which may be time consuming in the best case, or impossible to obtain in the worst case. Iterative methods can also be used for strain minimization \cite{buja2008data}, which allows the incorporation of weights and missing distances, at the expense of losing the global convergence guarantees. Due to the lack of geometric intuition behind it, strain minimization results in inferior embeddings \cite{elad2003bending} and has some other drawbacks. For example, classical scaling exhibits large sensitivity to outliers since the double centering transformation spreads the effect of a single outlier to the rest of the points in the same row/column.

\paragraph{Contribution.} Our work leverages the recent trend in geometry processing of using the truncated eigenbasis of the Laplace-Beltrami operator (LBO) to approximate smooth functions on manifolds. This favorable representation has been shown to improve classical scaling methods by orders of magnitude \cite{aflalo2013spectral}, as well as other key problems in geometry processing, e.g. non-rigid shape correspondence \cite{ovsjanikov2012functional}. 
By constraining the embedding to an affine subspace spanned by the leading eigenvectors of the Laplace-Beltrami operator, we show that LS-MDS exhibits a multi-resolution property, and can be solved significantly faster, making it applicable to larger size problems and to new applications. 

The rest of the article is arranged as follows: In Section \ref{sec:relworks} we review previous works and their relation to our work. In Section \ref{sec:LSMDSmin} we discuss the problem of stress minimization and describe in detail the SMACOF algorithm, a minimization-majorization type algorithm suited for this purpose. In Section \ref{sec:LBO} we give a short overview of the Laplace-Beltrami operator in the continuous and discrete settings, and the representation of band-limited signals on manifolds. In Section \ref{sec:subspaceMDS} we describe the core numerical scheme of our subspace LS-MDS approach, an algorithm we call \textit{spectral SMACOF} and a multiresolution approach. Finally, in Section \ref{sec:results} we present our results, followed by a concluding discussion in Section \ref{sec:conclusions}.

\section{Related work}\label{sec:relworks}
In this section we review previous work related to MDS, stress minimization techniques and subspace parametrization. 
The oldest version of MDS, called classical scaling, is due to Torgerson \cite{torgerson1952multidimensional}. 
The formulation of MDS as a minimization over a loss function is due to Shepard \cite{shepard1962analysis} and Kruskal \cite{kruskal1964multidimensional,kruskal1964nonmetric}. The SMACOF algorithm we describe in the next sections is due to de Leeuw  \cite{de1977applications}, but in fact was previously developed by Guttman \cite{guttman1968general}. For a comprehensive overview of MDS problems and algorithms we refer the reader to \cite{cox2000multidimensional}.
A variety of methods have been proposed to accelerate classical scaling over the years, trying to reduce the size of the input pairwise distance matrix.
\cite{de2003global} and \cite{brandes2006eigensolver} propose to embed only a subset of the points and find the rest by interpolation.
\cite{bengio2004out} uses the Nystr\"{o}m method to perform out of sample extension to an existing embedding. However, it does not make use of the geometry of the data. \cite{aflalo2013spectral} 
interpolate the distance matrix from a subset of pairwise distances by representing it in the Laplace-Beltrami eigenbasis and forcing it to be smooth on the manifold.
\cite{shamai2015classical} also uses the LBO and the Nystr\"{o}m method to interpolate the distance matrix from a set of known rows. This time the interpolation is formulated in the spatial domain so there is no need to compute a basis.

Regarding stress based scaling, \cite{bronstein2006multigrid} proposed a multi-grid approach. They rely heavily on the spatial representation of the shapes in switching between resolutions, which makes their scheme hard to generalize. \cite{rosman2008fast} used vector extrapolation techniques to accelerate the convergence rate of the SMACOF algorithm. This acceleration technique is a meta-algorithm that can be put on top of our algorithm as well. \cite{de1980multidimensional} considers MDS problems with linear restrictions on the configurations, essentially constraining the embedding to belong to a subspace. \cite{lawrence2011user} uses a subspace method for LS-MDS, but their subspace is an ad-hoc construction used for multi-channel image visualization. 

Linear subspace methods are common in machine learning and pattern recognition applications where they are used to restrict a signal to some low-rank subspace, usually an eigenspace of some matrix, which preserves its important details. Examples include principle component analysis (PCA), locally linear embedding (LLE), linear discriminant analysis (LDA) and many more \cite{saul2006spectral}.
One of those methods which is related to our algorithm is the Laplacian eigenmaps method \cite{belkin2003laplacian}. It is essentially an MDS problem using only local distances. Its outcome is an embedding of the distance matrix into the first $m$ eigenvectors of the Laplace-Beltrami operator. In some sense, our proposed approach is a merge between Laplacian eigenmaps and LS-MDS. A related line of works uses Laplacian (spectral) regularization in the context of low-rank subspace constructions on graphs \cite{shahid2015robust,kalofolias2014matrix}.

Subspace methods are also common in numerical optimization to accelerate and stabilize optimization algorithms. A review can be found here \cite{yuan2014review}. 
It is also worth mentioning the work of \cite{kovalsky2016accelerated} which uses the cotangent Laplacian as a proxy for the Hessian in various geometric optimization problems. Ours is a similar but different approach. We employ a subspace method which effectively projects the Hessian on the subspace of the leading eigenvectors of the Laplacian, thus solving a series of small dimensional problems, rather than approximate the Hessian using the Laplacian while retaining the dimension of the original problem.
\section{LS-MDS and stress minimization}\label{sec:LSMDSmin}
Given an $N\times N$ symmetric matrix $\bb{D}$ of dissimilarities, least squares MDS (LS-MDS) is an MDS model that looks for an Euclidean embedding $\bb{X}\in \mathbb{R}^{N\times m}$  by minimizing the stress function (\ref{eq:kruskalstress}). The latter can be minimized by nonlinear optimization techniques, e.g. gradient descent\footnote{Note also that (\ref{eq:kruskalstress}) is not differentiable everywhere.}. One particularly elegant and popular algorithm is SMACOF \cite{de1977applications,bronstein2008numerical} -- a majorization-minimization type algorithm
which can be shown equivalent to gradient descent with a specific choice of the step size \cite{bronstein2008numerical}. In light of that, we consider it as a representative for the class of first order methods for the minimization of (\ref{eq:kruskalstress}).
The idea is to use the Cauchy-Schwarz inequality to construct a \textit{majorizing function} $h(\bb{X},
\bb{Z})$ for the stress, obeying
$h(\bb{X},\bb{Z}) \geq \sigma(\bb{X})$ for every $\bb{X}$ and $\bb{Z}$ and $h(\bb{X},\bb{X}) =\sigma(\bb{X})$.
To do so, one should first note that (\ref{eq:kruskalstress}) can be written alternatively as
\begin{equation}\label{eq:stressAlternative}
\sigma(\bb{X})=\trace(\bb{X}^\Tr\bb{V}\bb{X})-2\trace(\bb{X}^\Tr\bb{B}(\bb{X})\bb{X})+\sum_{i<j}w_{ij}d_{ij}^2
\end{equation}
where $\bb{V}$ and $\bb{B}(\bb{X})$ are symmetric row-and column-centered matrices given by
\begin{equation}\label{eq:Vdef}
v_{ij}=
\begin{cases}
-w_{ij}\;\;&i\neq j\\
\sum_{k\neq i} w_{ik}\;\;&i=j
\end{cases}
\end{equation}

\begin{equation}\label{eq:Bdef}
b_{ij}=
\begin{cases}
-w_{ij}d_{ij}\|\bb{x}_i-\bb{x}_j\|^{-1}\;\;&i\neq j,\bb{x}_i\neq\bb{x}_j\\
0\;\;&i\neq j,\bb{x}_i=\bb{x}_j\\
-\sum_{k\neq i} b_{ik}\;\;& i=j,\\
\end{cases}
\end{equation}
and define
\begin{equation}\label{eq:majorizingFunction}
h(\bb{X},\bb{Z})=\trace(\bb{X}^\Tr\bb{V}\bb{X})-2\trace(\bb{Z}^\Tr\bb{B}(\bb{Z})\bb{X})+\sum_{i<j}w_{ij}d_{ij}^2,
\end{equation}
which is a convex (quadratic) function in $\bb{X}$.
Therefore, if we use the following iteration
\begin{align}\label{eq:MDSupdate}
\bb{X}_{k+1}&=\arg\!\min_{\bb{X}}h(\bb{X},\bb{X}_k) =\bb{V}^{\dagger}\bb{B}_k\bb{X}_k
= \left(\bb{V}+\frac{1}{N}\bb{1}\bb{1}^\Tr\right)^{-1}\bb{B}_k\bb{X}_k,
\end{align}
where $\bb{B}_k\equiv \bb{B}(\bb{X}_k)$ and $\bb{V}^{\dagger}$ denotes the pseudo-inverse of the rank deficient matrix $\bb{V}$, we obtain a monotonically decreasing sequence of stress values
\begin{equation}\label{eq:majorizationIneq}
\sigma(\bb{X}_{k+1})\leq h(\bb{X}_{k+1},\bb{X}_{k}) \leq h(\bb{X}_k,\bb{X}_{k})=\sigma(\bb{X}_k).
\end{equation}•
Equation (\ref{eq:majorizationIneq}) follows directly from (\ref{eq:majorizingFunction}) and (\ref{eq:MDSupdate}); in the last passage, we used the fact that
\begin{equation}\label{eq:pinvV}
\bb{V}^{\dagger} = \left(\bb{V}+\frac{1}{N}\bb{1}\bb{1}^\Tr\right)^{-1}-\frac{1}{N}\bb{1}\bb{1}^\Tr
\end{equation}
and that $\bb{1}^T\bb{B}_k=\bb{0}^T$.
The multiplicative update (\ref{eq:MDSupdate}) suggests that at each iteration, each coordinate in the current embedding is a weighted mean of the coordinates of the previous embedding, starting with an initial embedding $\bb{X}_0$, where the weights are given by the ratio of the pairwise distances, $d_{ij}\|\bb{x}_i-\bb{x}_j\|^{-1}$. 
In what follows, we refer to (\ref{eq:majorizationIneq}) as 
the \textit{majorization inequality}.

Each iteration of SMACOF requires the computation of the pairwise Euclidean distances between all points in the current embedding, a task of complexity $\mathcal{O}(N^2)$, and the solution of a linear system involving the pseudo-inverse of the matrix $\bb{V}$.  Despite being rank deficient (in fact, $\bb{V}$ is the graph Laplacian of the graph encoded by the adjacency matrix $\bb{W}$, i.e., the pairwise weights), $\bb{V}$ is not necessarily sparse, and computing its pseudo-inverse 
is of order $\mathcal{O}(N^3)$, except when $\bb{V}$ has a special form.
For example, in case all the weights $w_{ij}$ are equal to $1$, we get 
\begin{equation}
\bb{V}^{\dagger} = \frac{1}{N}\left(\bb{I}-\frac{1}{N}\bb{1}\bb{1}^\Tr \right),
\end{equation}•
and the SMACOF iteration reduces to
\begin{equation}\label{eq:unweightedMDSUpdate}
\bb{X}_{k+1}=\frac{1}{N}\bb{B}_k\bb{X}_k.
\end{equation}
To summarize, a single iteration of SMACOF consist of two steps: \emph{Majorization} -- the construction of $\bb{B}(\bb{X}_k)$ (\ref{eq:Bdef}), which requires $\mathcal{O}(N^2)$ non-linear operations; and \emph{Minimization} -- the solution of (\ref{eq:MDSupdate}), which in general requires $\mathcal{O}(N^3)$ for the factorization of $\bb{V}+\frac{1}{N}\bb{1}\bb{1}^\Tr$ in the first iteration, and $\mathcal{O}(N^2m)$ operations for solving the linear system using forward/backward substitution in the following iterations.


\section{Laplace-Beltrami operator and band-limited signals}\label{sec:LBO}
The Laplace-Beltrami operator is an analogue of the Euclidean Laplacian defined on a manifold, which has become a standard tool in the field of geometry processing.
For a closed compact manifold surface $S$, let $\Delta$ denote its
Laplace-Beltrami differential operator. Consider the equation $\Delta \phi = \lambda \phi$.
The pair $\phi,\lambda$ are an eigenpair of $\Delta$.
Note that $\lambda = 0$ is always an
eigenvalue, the corresponding eigenfunctions are constant
functions.
The eigenvalues of the Laplace-Beltrami operator are
non-negative and constitute a discrete set. We will assume
that the eigenvalues are distinct, so we can put them into ascending order, $0=\lambda_0 < \lambda_1 < \lambda_2 < \dots < \lambda_i < \ldots$. The appropriately normalized eigenfunction corresponding
to $\lambda_i$ will be denoted by $\phi_i$. The normalization is achieved
using the $L_2$ inner product. Given two functions $f$ and $g$ on
the surface, their inner product is denoted by $\left<f,g\right>$, and is
defined as the surface integral 
\begin{equation} 
\left<f,g\right> = \int_S fg \;da.
\end{equation}
Since the Laplace-Beltrami operator is Hermitian, the 
eigenfunctions corresponding to its different eigenvalues are
orthogonal. Given a function $f$ on the surface, one can
expand it in terms of the eigenfunctions 
$f = \sum_{i=1}^\infty \alpha_i\phi_i$, where the coefficients are $\alpha_i = \left<f,\phi_i\right>$. 
We refer to signals for which $\left<f,\phi_i\right> =0\; \forall i>p$ as $p$-bandlimited. This intuitively suggests that to smooth a function one should simply discard the coefficients corresponding to the larger eigenvalues, i.e. truncate the infinite expansion
above. A variety of discretizations exist for the Laplacian, each with its own pros and cons. We use the cotangent Laplacian of \cite{pinkall1993computing}. Instead of using the usual trigonometric construction, it can also be expressed in terms of the edge lengths and Heron's formula. See \cite{boscaini2014shape} for example.
The result is a generalized eigenvalue problem $\bb{W}\bb{\phi}=\lambda\bb{A}\bb{\phi}$ where $\bb{W}$ is a symmetric positive definite matrix called \textit{stiffness matrix} and $\bb{A}$ is a diagonal matrix called \textit{the mass matrix}.
Notice that in the discrete setting, this basis is orthonormal with respect to the inner product on the manifold $\left<\bb{f},\bb{g}\right>_{\bb{A}} = \bb{f}^\Tr\bb{A}\bb{g}$.

\section{Subspace LS-MDS}\label{sec:subspaceMDS}
The dependence of a single iteration of SMACOF on $N^2$ (or $N^3$) severely limits its applicability in embedding a large number of points. Deviating from the LS-stress model (\ref{eq:kruskalstress}) makes things even worse. Different stress models like the $L_1$ stress, which could in theory be solved as a series of weighted MDS (WMDS) problems using an iteratively reweighted least squares technique, require very long time to converge if one uses the regular SMACOF iterations. Moreover, adding linear constraints on the configuration would require the solution of a large quadratic program in each iteration, making it practically impossible to use. 

In order to reduce the dependence on $N$, we restrict the embedding to lie within a low-dimensional subspace of $\mathbb{R}^{N\times m}$. Low-rank representation of signals is a very popular approach in a variety of machine learning applications, and have also been applied previously to stress majorization \cite{lawrence2011user}. Our goal is to show that for multidimensional scaling problems arising in geometry processing, a particular choice of subspace, namely the subspace of band-limited signals on a manifold, enables to accelerate stress majorization by a significant amount. Unlike the ad-hoc construction in \cite{lawrence2011user}, this subspace is a geometric construction. We also draw a connection between the number of samples, the band-limit, and the quality of the approximation. Although only an observational result at this stage, similar results have been obtained for the case of reconstruction of band-limited signals on graphs \cite{puy2016random}.

We assume that an initial manifold $\mathcal{M}_0$ embedded in $\mathbb{R}^{m}$ that encodes the local relationship between points is easily obtainable, and denote the discrete samples of this manifold by $\bb{X}_0$.
The (discrete) embedding $\bb{X}$ can be written as $\bb{X}=\bb{X}_0+\bb{\delta}$, and we can reformulate MDS in terms of the displacement field $\bb{\delta}$. Our observation is that since the embedding found by LS-MDS maintains the local relationship between samples, as encouraged by the stress term (\ref{eq:kruskalstress}), $\bb{\delta}$ can be modeled as a band-limited signal (i.e. "smooth") on this manifold. We do that by explicitly representing the displacement field as a linear combination of the first $p$ eigenvectors of the Laplace-Beltrami operator on the manifold $\mathcal{M}_0$. We denote the concatenation of the first $p$ eigenvectors by $\bb{\Phi}\in\mathbb{R}^{N
\times p}$, and the problem therefore becomes
\begin{equation}\label{eq:subspaceStressMin}
\min_{\bb{\alpha}\in\mathbb{R}^{p\times m}}\sigma(\bb{X}_0+\bb{\Phi}\bb{\alpha})
\end{equation}
The new update step in terms of $\bb{\alpha}$ is
\begin{align}\label{eq:spectralSmacofUpdate}
\bb{\alpha}_{k+1}&=\arg\!\min_{\bb{\alpha}}h(\bb{X}_0+\bb{\Phi}\bb{\alpha},\bb{X}_k) \\
&= \left(\bb{\Phi}^\Tr\bb{V}\bb{\Phi}\right)^{\dagger}\bb{\Phi}^\Tr\left(\bb{B}_k\bb{X}_k-\bb{V}\bb{X}_0\right),
\end{align}
where $\bb{X}_{k+1}=\bb{X}_0+\bb{\Phi}\bb{\alpha}_{k+1}$.
Notice that the majorization inequality (\ref{eq:majorizationIneq}) still holds, since the requirement $\sigma(\bb{X}) \le h(\bb{X},\bb{Z})$ is true for any $\bb{X}$ and $\bb{Z}$, so it must also be true for $\bb{X}$ restricted to an affine subspace
\begin{equation}\label{eq:majorizationIneqExtended}
\sigma(\bb{X}_{k+1})\leq h(\bb{X}_0+\bb{\Phi}\bb{\alpha}_{k+1},\bb{X}_k) \leq \sigma(\bb{X}_k).
\end{equation}

Restricting the embedding to lie within a rank $p$ subspace reduces the cost of solving the linear system (\ref{eq:MDSupdate}) to
$\mathcal{O}(p^2m)$ (amortized $\mathcal{O}(p^3)$). This may have a profound effect on the \textit{minimization} step of SMACOF for weighted MDS problems or when additional terms are added to the stress. However, the main bottleneck remains in the \textit{majorization} step, which requires the computation of all pairwise Euclidean distances in the current embedding. Our main observation is that when the displacement field is $p$-bandlimited, i.e., can be written as a linear combination of the first $p$ Laplace-Beltrami eigenvectors, it is enough to use  $q=cp\ll N$
points to get an embedding which is almost as good, in terms of stress value, as the embedding achieved by using all the points. The quantity
$1<c=\frac{q}{p}\leq 2$ is a sampling ratio which can be found by experimentation. In practice, we set $c = 2$ in all our experiments. Denoting the set of sampled points by 
$\{s_1,\ldots,s_{q}\}\in \mathcal{S}\subset \{1,\ldots, N\},$
we define a sampling matrix $\bb{S}^{q\times N}$ with $\bb{S}_{ij}=1$ if $j=s_i$ and $0$ otherwise,
and instead of minimizing the objective of (\ref{eq:subspaceStressMin}), we minimize $\sigma(\bb{S}\bb{X}_0+\bb{S}\bb{\Phi}\bb{\alpha})$.
Each iteration of SMACOF now becomes
\begin{align}
{\bb{\alpha}}_{k+1}&=\arg\!\min_{\bb{\alpha}}h(\bb{S}\bb{X}_0+\bb{S}\bb{\Phi}\bb{\alpha},\bb{S}\bb{X}_k) \\
&= \left(\bb{\Phi}^\Tr\bb{S}^\Tr\bb{V}^\mathrm{s}\bb{S}\bb{\Phi}\right)^{\dagger}\bb{\Phi}^\Tr\bb{S}^\Tr\left({\bb{B}^\mathrm{s}}_k\bb{S}\bb{X}_k-\bb{V}^s\bb{S}\bb{X}_0\right) \label{eq:spectralSmacofUpdateSimplified}
\end{align}•
where $\bb{V}^\mathrm{s},\bb{B}^\mathrm{s}$ are the matrices defined in (\ref{eq:Vdef}),(\ref{eq:Bdef}) constructed only from the $q$ sampled points. We denote the solution and the final embedding by $\bb{\alpha}_*$ and $\bb{X}_* = \bb{X}_0+\bb{\Phi}\bb{\alpha}_*$, respectively, referring to the latter as \textit{spectral interpolation}.
We summarize this algorithm, dubbed \textit{spectral SMACOF}, in Alg. \ref{alg:spectralSMACOF}.


\begin{algorithm}[tb]
\SetArgSty{textnormal}
\DontPrintSemicolon
\KwIn{
\begin{itemize}
\item $\mathcal{M}$ - A discretized manifold (e.g. triangular mesh, polygonal mesh, point cloud).
\item  $\bb{W}$ - A matrix of pairwise weights.
\item $p$ - Number of Laplace-Beltrami eigenfunctions.
\item $q$ - Number of samples (default: $q =2p$).
\item $\mathcal{M}_0$ - A discretized initial manifold on which we can define a Laplace-Beltrami operator.
\item $\bb{X}_0 \in \mathbb{R}^{N\times m}$ - Initial embedding (discrete samples of $\mathcal{M}_0$).
\item $a_{\mbox{tol}}$ - Absolute tolerance.
\item $r_{\mbox{tol}}$ - Relative tolerance.
\item $\mbox{maxiter}$ - Maximum number of iterations.
\end{itemize}
}
\KwOut{Euclidean embedding $\bb{X}\in \mathbb{R}^{N\times m}$.\\}
\vspace{5pt}
$[\bb{D}\;,\;\bb{S}] = \mbox{FPS}(\mathcal{M})$ - 
Sample $\mathcal{M}$ with farthest point sampling and return sampling matrix $\bb{S}$ and matrix of geodesic distances between samples $\bb{D}$.\; 
 $[\bb{\Phi}\;,\bb{\Lambda}]=\mbox{LBO\_basis}(\mathcal{M}_0,p)$\ - Compute p leading eigenvectors $\bb{\Phi}$ and eigenvalues $\bb{\Lambda}$ of Laplace-Beltrami operator on $\mathcal{M}_0$.\;
Compute ${\bb{V}^\mathrm{s}}$ according to (\ref{eq:Vdef}).\;
Compute and store $(\bb{\Phi}^\Tr\bb{S}^\Tr\bb{V}^\mathrm{s}\bb{S}\bb{\Phi})^{\dagger}$ and $\bb{V}^\mathrm{s}\bb{S}\bb{X}_0$.\;
$k\gets 1$\;
\While{$ k\leq \mbox{maxiter} $ {\rm{\bf and}}  $\sigma(\bb{S}\bb{X}_k)>a_{\mbox{tol}} $ {\rm{\bf and}}  $\left(1-\frac{\sigma(\bb{S}\bb{X}_k)}{\sigma(\bb{S}\bb{X}_{k-1})}\right)>r_{\mbox{tol}}$}{
 Compute ${\bb{B}^\mathrm{s}}_k\equiv\bb{B}(\bb{S}\bb{X}_k)$ according to (\ref{eq:Bdef}).\;
  $\bb{\alpha}_{k+1} = \left(\bb{\Phi}^\Tr\bb{S}^\Tr\bb{V}^\mathrm{s}\bb{S}\bb{\Phi}\right)^{\dagger}\bb{\Phi}^\Tr\bb{S}^\Tr\left({\bb{B}^\mathrm{s}}_k\bb{S}\bb{X}_k-\bb{V}^\mathrm{s}\bb{S}\bb{X}_0\right)$\;
 $\bb{X}_{k+1} = \bb{X}_0+\bb{\Phi}\bb{\alpha}_{k+1}$\;
 $k \gets k + 1$\;
 }
\Return{$\bb{X}_{k+1}$.}\;
\caption{{\sc Spectral SMACOF}\label{alg:spectralSMACOF}}
\end{algorithm}

Notice that the pairwise distances are only required between the sampled points. In order 
to sample the points, we employ the \textit{farthest point sampling} strategy \cite{eldar1997farthest}, computing the distances as part of the sampling process.
Though the algorithm requires the computation of the truncated eigenbasis of the LBO, in order to apply (\ref{eq:spectralSmacofUpdateSimplified}) we do not need the full basis $\bb{\Phi}$, but only its sampled version $\bb{S}\bb{\Phi}$. The full basis is only needed for the final interpolation of the displacement field $\bb{\delta}$, which can also be reformulated as an optimization problem
\begin{equation}
\min_{\bb{\delta}} \|\bb{S}\bb{\delta}-\bb{S}\bb{\Phi}\bb{\alpha}_*\|_\mathrm{F}^2+\lambda\left<\bb{\delta},\bb{L}\bb{\delta}\right>,
\end{equation}
where $\bb{L}$ is the Laplacian matrix.
This opens up the possibility of reducing the amount of time required to construct the basis, and is an issue we leave for future research.

To summarize, a single iteration of \textit{spectral SMACOF} consist of two steps (excluding the interpolation which can be deferred to the final step):
\emph{Majorization} -- the construction of $\bb{B}^\mathrm{s}(\bb{X}_k)$, which requires $\mathcal{O}(q^2)$ operations; and
\emph{Minimization} -- the solution of (\ref{eq:spectralSmacofUpdateSimplified}), which in general requires $\mathcal{O}(p^3)$ for the computation of $(\bb{\Phi}^\Tr\bb{S}^\Tr\bb{V}^\mathrm{s}\bb{S}\bb{\Phi})^{\dagger}$ in the first iteration, and $\mathcal{O}(p^2m)$ operations in the succeeding iterations.

\subsection{Multi-resolution scheme for stress minimization}\label{subsec:MRMDS}
To control the trade off between convergence speed and approximation quality, we employ a multi-resolution scheme. 
 Multi-resolution and multi-grid methods have been previously employed for distance scaling with limited success \cite{bronstein2006multigrid}. Since those methods relied solely on the spatial representation of the embedding, the switch between resolution levels required cumbersome decimation and interpolation operators which were hard to generalize. The spectral formulation provides those operators practically "for free", where the decimation simply requires sampling the basis functions at the sampling points, and the interpolation is carried out using the full subspace basis.
 
The scheme consists of two distinct hierarchies, one for the spatial domain and one for the spectral domain. The two hierarchies are quantized and stored in the vectors
$\bb{q} = [q_1, \cdots, q_{r_q} ]^\Tr$ and
$\bb{p} = [p_1, \cdots, p_{r_p} ]^\Tr$,
where the last level for both could possibly be $N$, i.e. using the full set of vertices, in which case we get a ''pure'' subspace MDS, or using the complete basis, in which case we get an equivalent of the landmark MDS method \cite{de2003global}, or both, which reduces to applying a regular SMACOF iteration. It is important that each pair  $q_i,p_j$ complies with the ''sampling criterion'', $q_i\geq cp_j$.

We highlight two properties of this scheme: First, for a fixed $q$ and varying $p$, it follows from the majorization inequality that if we intialize each resolution level with the embedding from the previous resolution, we get a monotonic decrease in stress values. i.e., in the ''pure'' subspace setting $(q=N)$, spectral SMACOF produces a monotonically decreasing series of stress values, just like the regular SMACOF. 

Second, for a fixed $p$ with varying $q$, we observe that there is a jump in the stress value around $q\approx cp$, consistent with our observation in the derivation of the sampling criterion, and from there the approximation error and stress value decrease, though not monotonically. 

\section{Results}\label{sec:results}
In this section we compare the spectral SMACOF algorithm to other LS-MDS algorithms in terms of running time and quality of the embedding. We implemented the algorithms in Matlab\textsuperscript{\textregistered} using fast vector arithmetic. All the experiments were conducted on an Intel NUC6i7KYK equipped with Intel\textsuperscript{\textregistered}  Core\textsuperscript{TM} i7-6770HQ CPU 2.6GHz and 32 GB RAM. In all experiments we set $q=2p$, and use three resolution levels $\bb{q} =[200,600,N]$ and $\bb{p} = [100, 300, N]$.
For the spectral SMACOF algorithm we use the following parameters: $a_{\mbox{tol}}=0,\;r_{\mbox{tol} }=10^{-4},\;\mbox{maxiter}=100$  (see Alg. (\ref{alg:spectralSMACOF})), 
except for the full-resolution for which we set $r_{\mbox{tol} }=10^{-5}$.
For the SMACOF algorithm we use $r_{\mbox{tol} }=10^{-5},\;\mbox{maxiter}=5000$. We run it until it reaches the same value of stress achieved by the spectral SMACOF algorithm or lower.
We also compare to the algorithm proposed in \cite{rosman2008fast} which uses vector extrapolation methods (RRE) to accelerate SMACOF. Every $k$ iterations of regular SMACOF it tries to decrease the stress further by using a linear combination the previous descent directions. 
This acceleration technique is a meta-algorithm that can be put on top of the spectral SMACOF algorithm as well, and we only put it here for illustration purposes. In the following examples we set $k=5$.

\begin{figure}[t]
 
\subfloat[]{\label{fig:michael2conv}\includegraphics[width=0.493\textwidth]{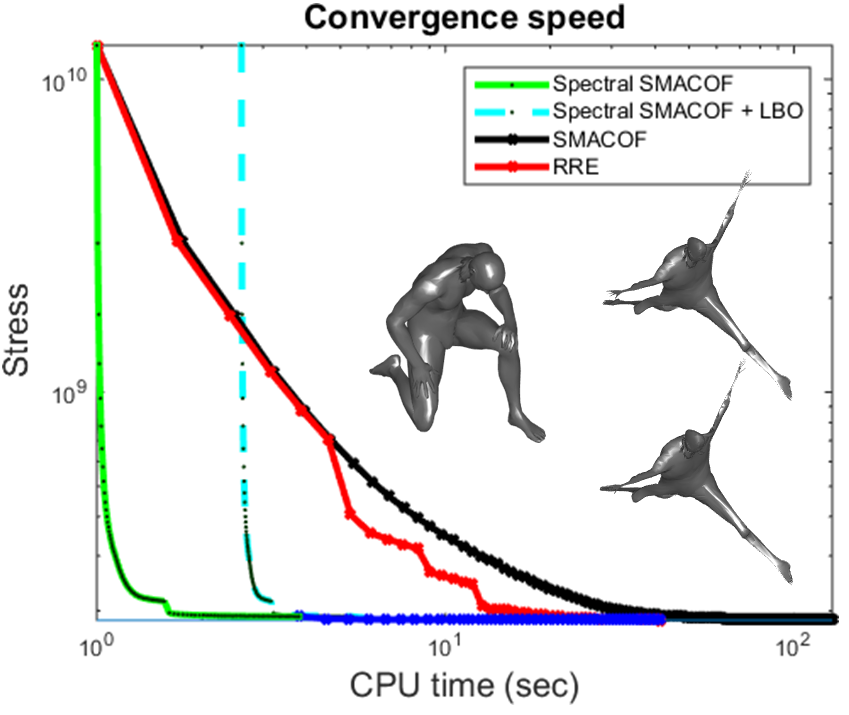}}%
 \enskip
\subfloat[]
{\label{fig:canoicalFormsWeighted_a}\includegraphics[width=0.493\textwidth]{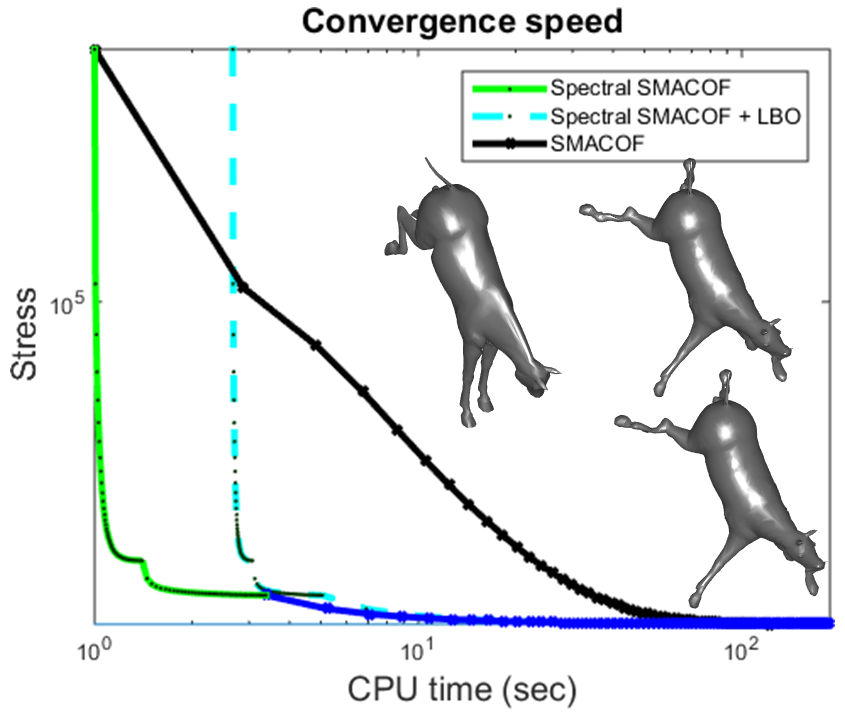}}

\caption[]{Convergence rate of Spectral SMACOF compared to regular SMACOF and RRE SMACOF \cite{rosman2008fast}. The dashed cyan plot is the accumulated time of spectral SMACOF and the subspace computation. Sub-figure \ref{fig:michael2conv} shows an unweighted MDS embedding.
Sub-figure \ref{fig:canoicalFormsWeighted_a} shows a weighted MDS embedding (relative stress). We can see a discontinuity when switching between resolution levels.
The dark blue plot shows the convergence of the last resolution level, which is practically regular SMACOF iterations. The left shape on each plot is the original shape which was also used to initialize the algorithms. The right shapes are the embeddings generated by spectral SMACOF (top) and regular SMACOF (bottom).
}
  \label{fig:canoicalForms}
\end{figure}

\paragraph{LS-MDS.} 
We apply spectral SMACOF to compute \textit{canonical forms} of several shapes from the low-res TOSCA dataset \cite{Tosca}, which have 3400 vertices. The canonical form is an isometric embedding of a surface $\bb{X}$ equipped with a metric $\bb{d}_X$, e.g. the geodesic metric, into $(\mathbb{R}^3,d_{\mathbb{R}^3})$. This embedding allows us to reduce the problem of non-rigid shape correspondence into a much simpler problem of rigid shape correspondence.
For more information see \cite{bronstein2008numerical}. 
Figure (\ref{fig:michael2conv}) shows running time for computing a canonical form using the unweighted stress. Figure (\ref{fig:canoicalFormsWeighted_a}) shows running time for a canonical form obtained with $w_{ij}=\sfrac{1}{d_{ij}^2}$. This kind of stress is commonly known as \textit{relative stress}. In both experiments we initialized with the original embedding.
The convergence rate of spectral SMACOF, measured in CPU time, is faster by two orders of magnitudes than the convergence rate obtained by regular SMACOF, even though the iteration count is much higher.
Since in the unweighted case the minimization step admits an especially simple form (see (\ref{eq:unweightedMDSUpdate})), this effect is more pronounced in the weighted case. Additional results and applications will appear in a longer version of this article.

\paragraph{Linearly constrained MDS.}
As another illustration, we use spectral SMACOF to perform image retargetting, the task of re-scaling an image in a non-uniform way such that the distortion will concentrate in non-salient regions. The saliency map is provided interactively by the user. We limit ourselves to horizontal re-scaling only, $\bb{\delta} = \bb{\Phi}\bb{\alpha}=\bb{\Phi}\left[\bb{\alpha}_1,\bb{0}\right]$. To do that we solve the following optimization problem
\begin{equation}
\min_{\bb{\alpha}_1} \sigma(\bb{X}_0+\bb{\Phi}\bb{\alpha}) +\mu \left<\bb{\alpha},\bb{L}\bb{\alpha}\right>\;\;\text{s.t.}\;\;\partial_x\bb{X}_0+\partial_x\bb{\Phi}\bb{\alpha}_1\geq \epsilon,
\end{equation}
i.e., we restrict the embedding to be monotonic
horizontally, and add a Dirichlet energy expressed in the Laplace-Beltrami basis. The linear constraints require the solution of a quadratic program in each iteration, instead of the update (\ref{eq:spectralSmacofUpdate}). Although the constraint has to be enforced at all points on the grid, we observed that enforcing the constraint only at the sampled points is enough, provided that the sampling is dense enough and that $\mu$ is large enough. The resulting quadratic program is pretty small, consisting of a few hundreds of variables and constraints, and we use Matlab's \texttt{quadprog} to solve it. We point out that even without the linear constraints, solving MDS problem of this size, consisting of millions of variables, is practically impossible with standard techniques.

\begin{figure}[t] 
  \centering
     \includegraphics[width=1\textwidth]{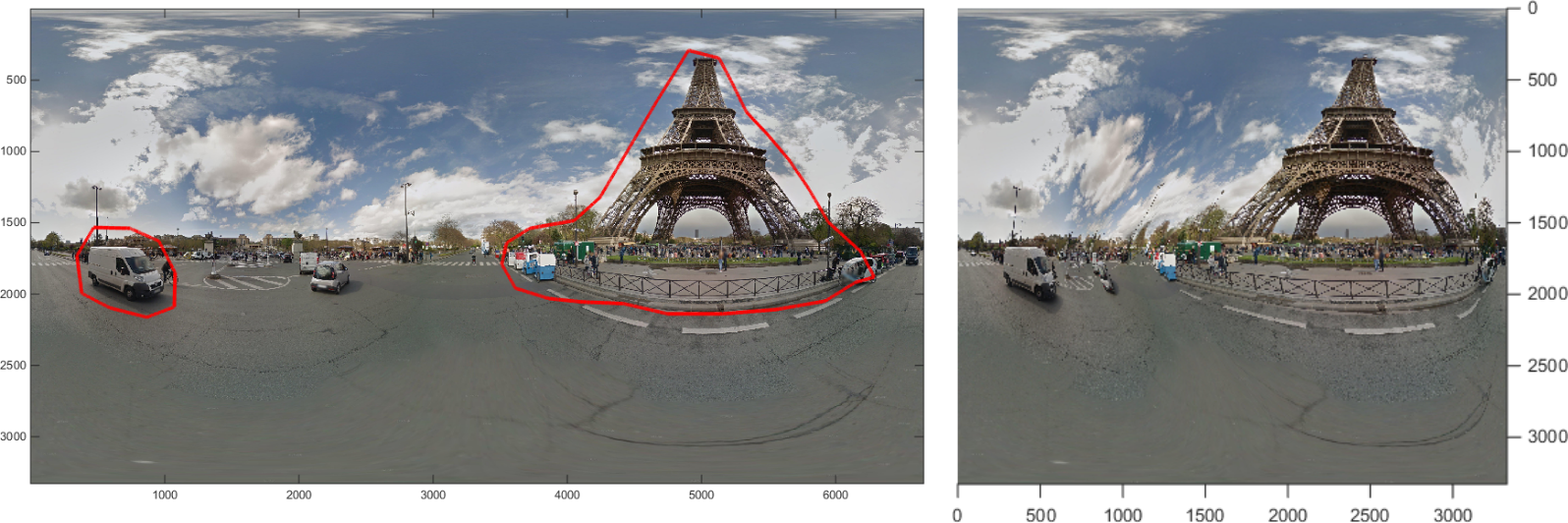}%


\caption[]{Left: Original panoramic image with salient regions circled in red . Right: Retargetted panorama, downscaled horizontally to half the size. Notice how the distortion is concentrated primarily in the non-salient regions.
  \label{fig:panorama}}
\end{figure}

Since the grid is Euclidean, the LBO basis is just the Fourier series, which can be computed in closed form. The initial manifold on which we construct the basis is the isotropically re-scaled image. We use sliding boundary conditions (i.e., Dirichlet on the horizontal boundaries and Neumann on the vertical). 
The results are presented in Figure (\ref{fig:panorama}). For this application we used the following parameters: $p=300$, $q=4p$, the image size is $3328\times 6656$, the deformation grid size is $832 \times 1664$. $\epsilon$ The size of the downscaled image is $3328\times 3328$, $\epsilon$ = half the horizontal spacing between deformation grid-cells.

\section{Conclusion}\label{sec:conclusions}
In this paper we showed that using a spectral representation for the embedding coordinates in LS-MDS problems arising in geometry processing uncovers a multi-resolution property of the numerical optimization scheme, which improves the convergence rate by orders of magnitude compared to other approaches, enabling its application to larger size problems and to more complex MDS models. The acceleration comes from two factors: First, we show experimentally that for various geometric optimization problems the displacement field between the final embedding and the initial embedding can be faithfully modeled as a bandlimited signal on the manifold, using only the first $p$ eigenvectors of the Laplacian, reducing the number of variables by a factor of $N/p$. Second, we observed that for such signals we can sample the spatial data by a rate of $\approx 2p/N$ and still converge to a local minimum of the full stress as if all the data was used. The quality of the embedding depends on the number of eigenvectors used and the number of points sampled, and changes from shape to shape. The relationship between $p$, $q$ and the approximation quality is only an observational result at this stage, and should be made more precise in future research.

In order to apply the method, the $p$ leading eigenvectors of the discrete Laplace-Beltrami operator have to be computed in advance. Though this is possible for moderately sized meshes and for specific larger size meshes with special structure for which a closed form expression for the eigenvectors exist, it remains the bottleneck in applying this method to very large meshes. A potential resolution to this bottleneck may come from the fact that the full basis vectors are only needed for the post process interpolation, which can be reformulated as an optimization problem that does not require the computation of the basis explicitly. We intend to explore this issue in future research.

\subsubsection{\ackname}\label{sec:ack}
This research was supported by the ERC StG grant no. 335491 and the ERC StG grant no. 307048 (COMET). Part of this research was carried out during a stay with Intel Perceptual Computing Group, Haifa, Israel.


\bibliography{mybib}{}

\begin{thebibliography}{10}
\providecommand{\url}[1]{\texttt{#1}}
\providecommand{\urlprefix}{URL }

\bibitem{Tosca}
Toolbox for surface comparison and analysis.
  \url{http://tosca.cs.technion.ac.il/book/resources_data.html}

\bibitem{aflalo2013spectral}
Aflalo, Y., Kimmel, R.: Spectral multidimensional scaling. Proceedings of the
  National Academy of Sciences  110(45),  18052--18057 (2013)

\bibitem{belkin2003laplacian}
Belkin, M., Niyogi, P.: Laplacian eigenmaps for dimensionality reduction and
  data representation. Neural computation  15(6),  1373--1396 (2003)

\bibitem{bengio2004out}
Bengio, Y., Paiement, J.f., Vincent, P., Delalleau, O., Roux, N.L., Ouimet, M.:
  Out-of-sample extensions for lle, isomap, mds, eigenmaps, and spectral
  clustering. In: Advances in Neural Information Processing Systems. pp.
  177--184 (2004)

\bibitem{borg2005modern}
Borg, I., Groenen, P.J.: Modern multidimensional scaling: Theory and
  applications. Springer Science \& Business Media (2005)

\bibitem{boscaini2014shape}
Boscaini, D., Eynard, D., Bronstein, M.M.: Shape-from-intrinsic operator. arXiv
  preprint arXiv:1406.1925  (2014)

\bibitem{brandes2006eigensolver}
Brandes, U., Pich, C.: Eigensolver methods for progressive multidimensional
  scaling of large data. In: International Symposium on Graph Drawing. pp.
  42--53. Springer (2006)

\bibitem{bronstein2008numerical}
Bronstein, A.M., Bronstein, M.M., Kimmel, R.: Numerical geometry of non-rigid
  shapes. Springer Science \& Business Media (2008)

\bibitem{bronstein2006multigrid}
Bronstein, M.M., Bronstein, A.M., Kimmel, R., Yavneh, I.: Multigrid
  multidimensional scaling. Numerical linear algebra with applications
  13(2-3),  149--171 (2006)

\bibitem{buja2008data}
Buja, A., Swayne, D.F., Littman, M.L., Dean, N., Hofmann, H., Chen, L.: Data
  visualization with multidimensional scaling. Journal of Computational and
  Graphical Statistics  17(2),  444--472 (2008)

\bibitem{cox2000multidimensional}
Cox, T.F., Cox, M.A.: Multidimensional scaling. CRC press (2000)

\bibitem{de1977applications}
De~Leeuw, J., Barra, I.J., Brodeau, F., Romier, G., Van~Cutsem, B., et~al.:
  Applications of convex analysis to multidimensional scaling. In: Recent
  Developments in Statistics. Citeseer (1977)

\bibitem{de1980multidimensional}
De~Leeuw, J., Heiser, W.J.: Multidimensional scaling with restrictions on the
  configuration. Multivariate analysis  5,  501--522 (1980)

\bibitem{de2003global}
De~Silva, V., Tenenbaum, J.B.: Global versus local methods in nonlinear
  dimensionality reduction. Advances in neural information processing systems
  pp. 721--728 (2003)

\bibitem{elad2003bending}
Elad, A., Kimmel, R.: On bending invariant signatures for surfaces. IEEE
  Transactions on pattern analysis and machine intelligence  25(10),
  1285--1295 (2003)

\bibitem{eldar1997farthest}
Eldar, Y., Lindenbaum, M., Porat, M., Zeevi, Y.Y.: The farthest point strategy
  for progressive image sampling. IEEE Transactions on Image Processing  6(9),
  1305--1315 (1997)

\bibitem{guttman1968general}
Guttman, L.: A general nonmetric technique for finding the smallest coordinate
  space for a configuration of points. Psychometrika  33(4),  469--506 (1968)

\bibitem{kalofolias2014matrix}
Kalofolias, V., Bresson, X., Bronstein, M., Vandergheynst, P.: Matrix
  completion on graphs. arXiv preprint arXiv:1408.1717  (2014)

\bibitem{kovalsky2016accelerated}
Kovalsky, S.Z., Galun, M., Lipman, Y.: Accelerated quadratic proxy for
  geometric optimization. ACM Transactions on Graphics (TOG)  35(4),  134
  (2016)

\bibitem{kruskal1964multidimensional}
Kruskal, J.B.: Multidimensional scaling by optimizing goodness of fit to a
  nonmetric hypothesis. Psychometrika  29(1),  1--27 (1964)

\bibitem{kruskal1964nonmetric}
Kruskal, J.B.: Nonmetric multidimensional scaling: a numerical method.
  Psychometrika  29(2),  115--129 (1964)

\bibitem{lawrence2011user}
Lawrence, J., Arietta, S., Kazhdan, M., Lepage, D., O'Hagan, C.: A
  user-assisted approach to visualizing multidimensional images. IEEE
  transactions on visualization and computer graphics  17(10),  1487--1498
  (2011)

\bibitem{ovsjanikov2012functional}
Ovsjanikov, M., Ben-Chen, M., Solomon, J., Butscher, A., Guibas, L.: Functional
  maps: a flexible representation of maps between shapes. ACM Transactions on
  Graphics (TOG)  31(4), ~30 (2012)

\bibitem{pinkall1993computing}
Pinkall, U., Polthier, K.: Computing discrete minimal surfaces and their
  conjugates. Experimental mathematics  2(1),  15--36 (1993)

\bibitem{puy2016random}
Puy, G., Tremblay, N., Gribonval, R., Vandergheynst, P.: Random sampling of
  bandlimited signals on graphs. Applied and Computational Harmonic Analysis
  (2016)

\bibitem{rosman2008fast}
Rosman, G., Bronstein, A.M., Bronstein, M.M., Sidi, A., Kimmel, R.: Fast
  multidimensional scaling using vector extrapolation. SIAM J. Sci. Comput  2
  (2008)

\bibitem{saul2006spectral}
Saul, L.K., Weinberger, K.Q., Ham, J.H., Sha, F., Lee, D.D.: Spectral methods
  for dimensionality reduction. Semisupervised learning pp. 293--308 (2006)

\bibitem{shahid2015robust}
Shahid, N., Kalofolias, V., Bresson, X., Bronstein, M., Vandergheynst, P.:
  Robust principal component analysis on graphs. In: Proceedings of the IEEE
  International Conference on Computer Vision. pp. 2812--2820 (2015)

\bibitem{shamai2015classical}
Shamai, G., Aflalo, Y., Zibulevsky, M., Kimmel, R.: Classical scaling
  revisited. In: Proceedings of the IEEE International Conference on Computer
  Vision. pp. 2255--2263 (2015)

\bibitem{shepard1962analysis}
Shepard, R.N.: The analysis of proximities: multidimensional scaling with an
  unknown distance function. i. Psychometrika  27(2),  125--140 (1962)

\bibitem{torgerson1952multidimensional}
Torgerson, W.S.: Multidimensional scaling: I. theory and method. Psychometrika
  17(4),  401--419 (1952)

\bibitem{yuan2014review}
Yuan, Y.x.: A review on subspace methods for nonlinear optimization. In:
  Proceedings of the International Congress of Mathematics. pp. 807--827 (2014)

\end{thebibliography}
\bibliographystyle{splncs03}

\end{document}